\documentclass[conference]{IEEEtran}
\usepackage{charter}
\usepackage{mathrsfs}
\usepackage{latexsym}
\usepackage{pifont}
\usepackage{epsfig}
\usepackage{array}
\usepackage{url}
\usepackage{amssymb}
\usepackage{amsthm}
\usepackage{amsmath}
\usepackage{amsfonts}
\usepackage{statex}
\usepackage{mathrsfs}
\usepackage{graphicx}
\usepackage{float}
\usepackage{hyperref}
\usepackage{verbatim}
\usepackage{footnote}
\makesavenoteenv{environmentname}

\def\Pr{{\mathbb P}}

\long\def\symbolfootnote[#1]#2{\begingroup
\def\thefootnote{\fnsymbol{footnote}}\footnote[#1]{#2}\endgroup}

\ifCLASSINFOpdf
\else
\fi

\begin{document}

\newtheorem{theorem}{Theorem}
\newtheorem{lemma}{Lemma}
\newtheorem{proposition}{Proposition}
\newtheorem{corollary}{Corollary}[theorem]
\newtheorem{definition}{Definition}
\newtheorem{remark}{Remark}[theorem]
\newtheorem{notation}{Notation}
\newtheorem{example}{Example}

\title{
Delay Asymptotics with Retransmissions and Incremental Redundancy Codes over Erasure Channels
}

\author{\IEEEauthorblockN{Yang Yang\IEEEauthorrefmark{2}, Jian Tan\IEEEauthorrefmark{1}, Ness B. Shroff\IEEEauthorrefmark{2},
Hesham El Gamal\IEEEauthorrefmark{2}
}
\IEEEauthorblockA{\IEEEauthorrefmark{2}Department of Electrical and Computer Engineering\\
The Ohio State University, Columbus 43210, OH}
\IEEEauthorblockA{\IEEEauthorrefmark{1}IBM T.J. Watson Research, Hawthorne 10532, NY}
\vspace{-0.4cm}}
\maketitle

\begin{abstract}
Recent studies {\color{black} have shown} that retransmissions can cause heavy-tailed transmission delays even when packet sizes are light-tailed. Moreover, the impact of heavy-tailed delays persists even when packets size are upper bounded. The key question we study in this paper is how the use of coding techniques to transmit information, together with different system configurations, would affect the distribution of delay. To investigate this problem, we model the underlying channel as a Markov modulated binary erasure channel, where transmitted bits are either received successfully or erased. Erasure codes are used to encode information prior to transmission, which ensures that a fixed fraction of the bits in the codeword can lead to successful decoding. We {\color{black}use} incremental redundancy codes, where the codeword is divided into codeword trunks and these trunks are transmitted one at a time to provide incremental redundancies to the receiver until the information is recovered. We characterize the distribution of delay under two different scenarios: (I) Decoder uses memory to cache all previously successfully received bits. (II) Decoder does not use memory, where received bits are discarded if the corresponding information {\color{black}cannot} be decoded. 
In both cases, we consider codeword length with infinite and finite support. 
From a theoretical perspective, our results provide a benchmark to quantify the tradeoff between system complexity and the distribution of delay.

\end{abstract}

\IEEEpeerreviewmaketitle

\section{Introduction}
Retransmission is {\color{black}the} basic component {\color{black}used
  in} most medium access control protocols and it is used to ensure
reliable transfer of data over communication channels with failures
\cite{BG92}. Recent studies
\cite{HeavyTail}\cite{Jelen07ALOHA}\cite{Jelen07e2e} {\color{black}have
  revealed the surprising result that} retransmission-based protocols
could cause heavy-tailed transmission delays even if the packet length
is light tail distributed, resulting in {\it very long delays} and
possibly {\it zero throughput}. Moreover, \cite{TanShroff10} shows
that even when the packet sizes are upper bounded, the distribution of
delay, although eventually light-tailed, may still have a heavy-tailed
main body, and {\color{black}that} the heavy-tailed main body could
dominate even for relatively small values of the maximum packet
size. In this paper we investigate the use of coding techniques to
transmit information in order to alleviate the impact of heavy tails,
and substantially reduce the incurred transmission delay.


In our analysis, we focus on the {\it Binary Erasure
 Channel}. Erasures in communication systems can arise in different layers. At
the physical layer, if the received signal falls outside acceptable
bounds, it is declared as an erasure. At the data link layer, some
packets may be dropped because of checksum errors. At the network
layer, packets that traverse through the network may be dropped
because of buffer overflow at intermediate nodes and therefore never
reach the destination. All these errors can result in erasures in the
received bit stream.

In order to investigate how different coding techniques would affect the delay distribution, we {\color{black}use a} general coding framework called {\it incremental redundancy codes}. In this framework, each codeword is split into several pieces with equal size, which {\color{black}are called} codeword trunks. The sender sends only one codeword trunk at a time. If the receiver cannot decode the information, it will request the sender to send another piece of {\color{black}the} codeword trunk. Therefore, at every transmission, the receiver gains extra information, which is called incremental redundancy. 

In order to combat channel erasures, we use erasure codes {\color{black}as channel coding} to encode the information. Erasure codes represent a group of coding schemes which ensure that even when some portions of the codeword are lost, it is still possible for the receiver to recover the corresponding information. Roughly speaking, the encoder transforms a data packet of $l$ symbols into a longer codeword of $l_c$ symbols, where the ratio $\beta=l/l_c$ is called the code-rate. An erasure code is said to be near optimal if it requires slightly more than $l$ symbols, say $(1+\varepsilon)l$ symbols, to recover the information, where $\varepsilon$ can be made arbitrary small at the cost of increased encoding and decoding complexity. Many elegant low complexity erasure codes have been designed for erasure channels, e.g., Tornado Code \cite{TornadoCode}, LT code \cite{LT-code}, and Raptor code \cite{TON}. For the sake of simplicity, throughout the paper, we assume $\varepsilon=0$. In other words, any $\beta$ fraction of the codeword can recover the corresponding information {\color{black} and a lower $\beta$ indicates a larger redundancy in the codeword.}

\begin{figure}[h]
\begin{center}
\includegraphics [width=3.65in]{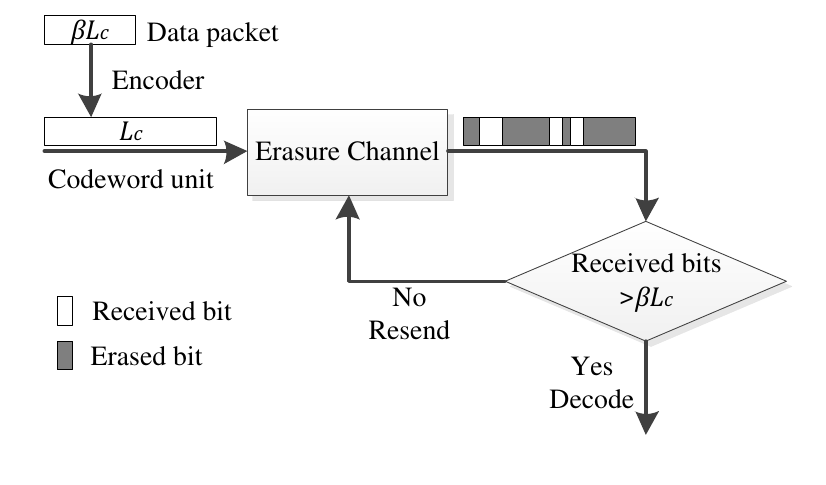}
\caption{Decoder that does not use memory scenario}
\label{erasurechannel}
\end{center}
\end{figure}

{\color{black}We specify different scenarios in this paper. In the
  first scenario, as shown in Fig. \ref{erasurechannel}, the entire
  codeword is transmitted as a unit, and received bits are simply
  discarded if the corresponding information cannot be recovered. Note
  that in this scenario, the decoder memory is not exploited for caching
  received bits across different transmissions. This scenario occurs
  because the receiver may not have the requisite computation/storage
  power to keep track of all the erasure positions and the bits that
  have been previously received, especially when the receiver is
  responsible for handling a large number of flows simultaneously. In
  the second scenario, we assume that the receiver has enough memory
  space and computational power to accumulate received bits from
  different (re)transmissions, which enables the use of incremental
  redundancy codes, where a codeword of length $l_c$ is split into $r$
  codeword trunks with equal size, and these codeword trunks are
  transmitted one at a time. At the receiver, all successfully
  received bits from every transmission are buffered at the receiver
  memory according to their positions in the codeword. If the receiver
  cannot decode the corresponding information, it will request the
  sender to send another piece of codeword trunk. At the sender, these
  codeword trunks are transmitted in a round-robin manner. We call
  these two scenarios {\bf Decoder that does not use memory} and {\bf
    Decoder that uses momery}, respectively.}


Given the above two different types of decoder, there are two more factors that can affect the distribution of delay. (I) {\bf Channel Dynamics:} In order to capture the time correlation nature of {\color{black}the} wireless channels, we assume that the channel is Markovian modulated. More specifically, we assume a time slotted system where one bit can be transmitted per time slot, and the current channel state distribution depends on channel states in the previous $k$ time slots. When $k=0$, it corresponds to {\color{black}the} {\it i.i.d.} channel model. (II) {\bf Codeword length distribution:} We assume throughout the paper that the codeword length is light tail distributed, which implies that the system works in a benign environment. We consider two different codeword length distributions, namely, codeword length with {\it infinite support} and codeword length with {\it finite support}, respectively. For the former, the codeword length distribution has an exponentially decaying tail with decay rate $\lambda$, for the latter, the codeword length has an upper bound $b$.

\subsection*{Contribution}
{\color{black}
The main contribution of this work is the following:
\begin{itemize}
\item
When decoder memory is not exploited, the tail of the delay distribution depends on the code rate. Specifically, we show that when the coding rate is above a certain threshold, the delay distribution is heavy tailed, otherwise it is light tailed.
This shows that substantial gains in delay can be achieved
over the standard retransmission case (repetition coding) by adding a certain amount of redundancy in the codeword. As mentioned earlier, prior work has shown that repetition coding results in heavy tailed delays even when the packet size are light tailed.
\item
When decoder memory is exploited, the tail of the delay distribution is always light-tailed. This implies that the use of receiver memory results in a further substantial reduction in the transmission delay.
\item
The aforementioned results are for the case when the codeword size can have infinite support. We also characterize the transmission delay for each of the above cases when the codeword size has finite support (zero-tailed), and show similar tradeoffs between the coding rate and use of receiver memory in terms of the main body of the delay distribution (rather than the eventual tail).
\end{itemize}}



The remainder of this paper is structured as follows: In Section \ref{sysmod}, we describe the system model. In Section \ref{uses} we consider the scenario where the decoder memory is exploited. Then, in Section \ref{doesnotuse} we investigate the situation where the decoder does not use memory. Finally, in Section \ref{sims}, we provide numerical studies to verify our main results.

\section{System Model}\label{sysmod}
The channel dynamics are modeled as a slotted system where one bit can be transmitted per slot. Furthermore, we assume that the slotted channel is characterized by a binary stochastic process $\{X_n\}_{n\geq1}$, where $X_n=1$ corresponds to the situation when the bit transmitted at time slot $n$ is successfully received, and $X_n=0$ when the bit is erasured (called an erasure). 

Since, in practice, the channel dynamics are often temporarily correlated, we investigate the situation {\color{black}in which} the current channel state distribution depends on the channel states in the preceding $k$ time slots. More precisely, for $\mathcal{F}_{n}=\{X_{i}\}_{i\leq n}$ and fixed $k$, we define $\mathcal{H}_{n}=\{X_{n},\ldots,X_{n-k+1}\}$ for $n\geq k\geq 1$ with $\mathcal{H}_{n}=\{\varnothing,\Omega\}$ for $k=0$, and assume that $\Pr[X_{n}=1|\mathcal{F}_{n-1}]=\Pr[X_{n}=1|\mathcal{H}_{n-1}]$ for all $n\geq k$. To put it another way, the augmented state $Y_n\triangleq[X_n,\ldots,X_{n-k}],n\geq k$ forms a Markov chain. Let $\Pi$ denote the transition matrix of the Markov chain $\left\{Y_n\right\}_{n\geq k+1}$, where
	\begin{align}
	\Pi=[\pi(s,u)]_{s,u\in\{0,1\}^k},\notag
	\end{align}
with $\pi(s,u)$ being the one-step transition probability from state $s$ to state $u$. Throughout this paper, we assume that $\Pi$ is irreducible and aperiodic, which ensures that this Markov chain is ergodic \cite{MarkovChain}. Therefore, for any initial value $\mathcal{H}_{k}$, the parameter $\gamma$ is well defined and given by
\begin{align}
\gamma=\lim_{n\to\infty}\Pr[X_n=1],\notag
\end{align}
and, from ergodic theorem (see Theorem 1.10.2 in \cite{MarkovChain})
\begin{align}
\Pr\left[\lim_{n\to\infty}\frac{\sum_{k=1}^nX_i}{n}=\gamma\right]=1,\notag
\end{align}
which means the long-term fraction of the bits that can be successfully received is equal to $\gamma$. Therefore, we call $\gamma$ the {\it channel capacity}.

In the degenerated case when $k=0$, we have a memoryless binary erasure channel (i.i.d. binary erasure channel). Correspondingly, $\mathcal{H}_n=\{\varnothing,\Omega\}$ and $\Pi=[\gamma]$. 

{\color{black}As mentioned in the introduction, we} study two different scenarios in this paper, namely {\it decoder that uses memory} and {\it decoder that does not use memory}. In the first scenario, the sender splits a codeword into $r$ codeword trunks with equal size and transmits them one at a time in a round-robin manner, while the receiver uses memory to cache all previously successfully received bits according to their positions in the codeword. In the second scenario, the receiver discards any successfully received bits if they cannot recover the corresponding information, and the sender transmits the entire codeword as a unit. 

We {\color{black}let $L_c$} denote the number of bits in the codeword with infinite support, and assume that there exist $\lambda>0$ and $z>0$ such that
\begin{align}
\lim_{x\to\infty}\frac{\log\Pr[x<L_c<x+z]}{x}=-\lambda.\label{Lcdef}
\end{align}
We {\color{black}let $L_c(b)$} denote the number of bits in the codeword with finite support, with $b$ being the maximum codeword length, and let $\Pr[L_c(b)>x]=\Pr[L_c>x|L_c<b]$ for any $x>0$.
We focus on erasure codes, where a fixed fraction ($0<\beta<1$) of bits in the codeword can lead to a successful decoding. We call this fraction $\beta$ code-rate.

Formal definitions of the number of retransmissions and the delays are given as follows:
\begin{definition}[Decoder that uses memory]\label{Nm}
The total number of transmissions for a codeword with variable length $L_c$ and number of codeword trunks $r$ when the decoder uses memory is defined as
	\begin{align}
	N_m^{(r)}\triangleq\inf&{\Bigg\{}n:\sum_{l=1}^r\sum_{i=(L_c/r)(l-1)+1}^{(Lc/r)l}\notag\\
	&\left.{\bold 1}\left(\sum_{j=1}^{\left\lfloor(n-l)/r\right\rfloor+1}X_{(j-1)L_c+i}\geq1\right)>\beta L_c\right\}.\notag
	\end{align}
The transmission delay is defined as $T_m^{(r)}=N_m^{(r)}L_c/r$. 
\end{definition}

\begin{definition}[Decoder that does not use memory]\label{Nf}
	The total number of transmissions for a codeword with variable length $L_c$ when the decoder does not use memory is defined as
	\begin{align}
	N_f\triangleq\inf\left\{n:\sum_{i=1}^{L_c}X_{(n-1)L_c+i}>\beta L_c\right\}.\notag
	\end{align}
	The transmission delay is defined as $T_f=N_fL_c$.
\end{definition}
	For a codeword with variable length $L_c(b)$, the corresponding numbers of transmissions and delays are denoted as $N_m^{(r)}(b),T_m^{(r)}(b),N_f(b),$ and $T_f(b)$, respectively.

\subsection*{Notations}
In order to present the main results, we introduce some necessary notations here. 
\begin{notation}
Let $\rho(M)$ denote the Perron-Frobenius eigen-value (see Theorem 3.11 in \cite{Dembozeitouni}) of the matrix $M$, which is the largest eigenvalue of $M$.
\end{notation}
\begin{notation}
For $k\geq 1$, {\color{black}let $\{s_i\}_{1\leq i\leq2^k}=\{0,1\}^k$} denote the state space of $\{Y_n\}_{n\geq k+1}$, where $s_i=[s_{i1},s_{i2},\ldots,s_{ik}]$ and $s_{ij}\in\{0,1\}$ $\forall i,j$. Then, we define a mapping $f$ from $\{s_i\}_{1\leq i\leq2^k}$ to $\{0,1\}$ as
\begin{align}
f(s_i)=1-s_{ik}.\notag
\end{align}
\end{notation}
\begin{notation}\label{RateFunction}
Let $\Lambda_n(\beta,\Pi)$ denote {\color{black}the} large deviation rate function, which is given by
	\begin{align}		
	 \Lambda_n\left(\beta,\Pi\right)&=\sup_\theta\left\{\theta(1-\beta)-\log\rho_n(\theta,\Pi)\right\},\notag
    \end{align}
    where\footnotemark
    \begin{align}
&\rho_n(\theta,\Pi)=\left\{
\begin{array}{ll}
\rho\left(e^{\theta D^{\otimes n}}\Pi^{\otimes n}\right)&k\geq1\\
(1-\gamma)^n+\left(1-(1-\gamma)^n\right) e^{\theta}&k=0
\end{array}
\right.\notag,\\
&D=\diag\left[f(s_1),f(s_2),\ldots,f(s_{2^k})\right]\text{ for }k\geq1\notag.
	\end{align}
\end{notation}
\footnotetext{For a matrix $A$, $A^{\otimes n}$ is the $n$-fold Kronecker product of $A$ with itself, or we can call it the $n^{th}$ Kronecker power of $A$.} 
\begin{notation}
Let $\mu_n$ denote the root of the rate function $\Lambda_n(\beta,\Pi)$. More precisely,
\begin{align}
\Lambda_n(\mu_n,\Pi)=0.\notag
\end{align}
\end{notation}
\begin{notation}\label{lambdas}
\begin{align}
	\alpha&=\inf\left\{n:\mu_n\geq\beta\right\}\notag\\
	\Lambda_1^o&=\inf_{n\in\mathbb{N}}\frac{\lambda+\Lambda_n(\beta,\Pi){\bf 1}(n\geq\alpha)}{n+1}\notag\\
	\Lambda_2^o&=\inf_{n\in\mathbb{N}}\frac{\lambda+\Lambda_{n+1}(\beta,\Pi){\bf 1}(n\geq\alpha-1)}{n+1}\notag\\
	\Lambda_3^o&=\left\{
	\begin{array}{ll}
	\lambda&\text{ if }\beta>\gamma\\
	\frac{\lambda r}{\lceil r\beta/\gamma\rceil}&\text{ if }\beta\leq\gamma
	\end{array}
\right..\notag
\end{align}
\end{notation}

\section{Decoder that uses Memory}\label{uses}
When the decoder uses memory to cache all previously successfully received bits, we can apply incremental redundancy codes, where the sender splits a codeword into $r$ codeword trunks and transmits one codeword trunk at a time. If the receiver, after receiving a codeword trunk, is not able to decode the corresponding information, it will use memory to cache the successfully received bits in the codeword trunk and request the sender to send another codeword trunk. In this way, at every transmission, the receiver gains extra information, which we call {\it incremental redundancy}. The sender will send these codeword trunks in a round-robin manner, meaning that if all of the codeword trunks have been requested, it will start over again with the first codeword trunk. It should be noted that incremental redundancy code is a fairly general framework in that if $r=1$, it degenerates to a fixed rate erasure code, while as $r$ approaches infinity, it resembles a rateless erasure code. 

\subsection{Codeword with infinite support}
When the distribution of codeword length $L_c$ has an exponentially decaying tail with decay rate $\lambda$, as indicated by Equation (\ref{Lcdef}), we find that the delay will always be light-tailed, and we characterize the decay rate in Theorem \ref{irc}.
\begin{theorem}\label{irc}
In the case when the decoder uses memory, when we apply incremental redundancy code with parameter $r$ to transmit codeword with variable length $L_c$, we obtain a lower and upper bound on the decay rate of delay,
\begin{align}
&-\liminf_{t\to\infty}\frac{\log\Pr\left[T_m^{(r)}>t\right]}{t}\leq\min\{\Lambda_2^o,\Lambda_3^o\},\notag\\
&-\limsup_{t\to\infty}\frac{\log\Pr\left[T_m^{(r)}>t\right]}{t}\geq\min\{\Lambda_1^o,\Lambda_3^o\}.\notag
\end{align}
In the special case when $r=1$,
\begin{align}
-\lim_{n\to\infty}\frac{\log\Pr\left[T_m^{(1)}>t\right]}{t}=\min\{\Lambda_1^o,\lambda\}.\notag
\end{align}
The definitions of $\Lambda_1^o, \Lambda_2^o$ and $\Lambda_3^o$ can be found in Notation \ref{lambdas}.
\end{theorem}
\begin{IEEEproof}
see Section \ref{ircproof}.
\end{IEEEproof}
\begin{remark}{\rm
From the definitions of $\Lambda_1^o, \Lambda_2^o$ and $\Lambda_3^o$ in Notation \ref{lambdas} we observe that firstly, the decay rate of delay when $r=1$ is no greater than the decay rate of delay when $r>1$ ($\min\{\Lambda_1^o,\lambda\}\leq\min\{\Lambda_1^o,\Lambda_3^o\}$), which means that incremental redundancy code ($r>1$) outperforms fixed rate erasure code ($r=1$); secondly, the decay rate of delay increases with the increase of $r$, which means we can reduce delay by increasing the number of codeword trunks $r$. These observations are verified through Example 1 in Section \ref{sims}.
}\end{remark}

\subsection{Codeword with finite support}
In practice, codeword length is bounded by the maximum transmission unit (MTU). Therefore, we investigate the case when the codeword has variable length $L_c(b)$, with $b$ being the maximum codeword length, and characterize the corresponding delay distribution in Theorem \ref{fse}.
\begin{theorem}\label{fse}
In the case when decoder uses memory, when we apply incremental redundancy code with parameter $r$ to transmit codeword with variable length $L_c(b)$, we get\\
1) for any $\eta>0$ and any $b_0>0$, we can find $b(\eta)>0$ such that for any $b>b(\eta)$, we have
$\forall t\in[n^o_2(b-b_0),n^o_2b]$,
	\begin{align}
		(1-\eta)\Lambda^b_1\leq-\frac{\log\Pr\left[T_m^{(r)}(b)>t\right]}{t}\leq(1+\eta)\Lambda^b_2.\notag
	\end{align}
2) in the special case when $r=1$, for any $\eta>0$ and any $b_0>0$, we can find $b(\eta)>0$ such that for any $b>b(\eta)$, we have
$\forall t\in[n^o_1(b-b_0),n^o_1b]$,
	\begin{align}
		(1-\eta)\Lambda^b\leq-\frac{\log\Pr\left[T_m^{(1)}(b)>t\right]}{t}\leq(1+\eta)\Lambda^b,\notag
	\end{align}
where
\begin{align}
n^o_1&=\arg\inf_{n\in\mathbb{N}}(\lambda+\Lambda_{n}(\beta,\Pi){\bf 1}(n\geq\alpha))/(n+1),\notag\\
	n^o_2&=\arg\inf_{n\in\mathbb{N}}(\lambda+\Lambda_{n+1}(\beta,\Pi){\bf 1}(n\geq\alpha-1))/(n+1),\notag\\
	\Lambda^b_1&=\Lambda_1^o+\min\{0,\Lambda_3^o-\Lambda_1^o\}{\bf1}(n_2^o=1),
	\notag\\	
	\Lambda^b_2&=\Lambda_2^o+\min\{0,\Lambda_3^o-\Lambda_2^o\}{\bf1}(n_2^o=1),
	\notag\\	
	\Lambda^b&=\Lambda_1^o+\min\{0,\lambda-\Lambda_1^o\}{\bf1}(n_1^o=1).
	\notag
\end{align}
\end{theorem}
\begin{IEEEproof}
see Section \ref{fseproof}.
\end{IEEEproof}
\begin{remark}{\rm This theorem shows that even if the codeword length has an upper bound $b$, the distribution of delay still has a light-tailed main body whose decay rate is similar as the decay rate of the infinite support scenario. The waist of this main body is $n_2^ob$ when $r>1$ and $n_1^ob$ when $r=1$. Since both $n_2^o$ and $n_1^o$ are independent of $b$, we know that the waist of this light-tailed main body scales linearly with respect to the maximum codeword length $b$. This theorem is verified through Example 2 in Section \ref{sims}.
}\end{remark}

\section{Decoder that does not use Memory}\label{doesnotuse}
For receivers that do not have the required computation/storage power, it is difficult to keep track of all the erasure positions and the bits that have been successfully received. Therefore, in this section, we study the case when the decoder does not use memory, {\color{black}as illustrated in Fig. \ref{erasurechannel}}. In this situation, since the receiver simply discards any successfully received bits if they cannot recover the corresponding information, it is better for the sender to transmit the whole codeword as a unit instead of dividing the codeword into pieces before transmission.

\subsection{Codeword with infinite support}

Interestingly, we observe an intriguing threshold phenomenon. We show that when the codeword length distribution is light-tailed and has an infinite support, the transmission delay is light-tailed (exponential) only if $\gamma>\beta$, and heavy-tailed (power law) if $\gamma<\beta$.

\begin{theorem}[Threshold phenomenon]\label{ptp}
In the case when decoder does not use memory and the codeword has variable length $L_c$, we get
\begin{enumerate}
\item if $\beta>\gamma$, then
	\begin{align}
		\lim_{n\to\infty}&\frac{\log\Pr\left[N_f>n\right]}{\log n}.\notag\\
		&=\lim_{t\to\infty}\frac{\log\Pr\left[T_f>t\right]}{\log t}=-\frac{\lambda}{\Lambda_1(\beta,\Pi)}.\notag
	\end{align}
\item if $\beta<\gamma$, then
	\begin{align}
		 \lim_{t\to\infty}\frac{\log\Pr\left[T_f>t\right]}{t}=-\min\left\{\lambda,\Lambda_1(\beta,\Pi)\right\}.\notag
	\end{align}
\end{enumerate}
The definition of $\Lambda_1(\beta,\Pi)$ can be found in Notation \ref{RateFunction}.
\end{theorem}
\begin{IEEEproof}
see Section \ref{ptpproof}.
\end{IEEEproof}
\begin{remark}{\rm
The tail distribution of the transmission delay changes from power law to exponential, depending on the relationship between code-rate $\beta$ and channel capacity $\gamma$. If $\lambda/\Lambda_1(\beta,\Pi)<1$, the system even has a zero throughput. 
}\end{remark}

\subsection{Codeword with finite support}
Under the heavy-tailed delay case when $\beta>\gamma$, we can further show that if the codeword length is upper bounded, the delay distribution still has a heavy-tailed main body, although it eventually becomes light-tailed.
\begin{theorem}\label{fsenm}
In the case when decoder does not use memory and the codeword has variable length $L_c(b)$, if $\beta>\gamma$, for any $\eta>0$, we can find $n(\eta)>0$ and $b(\eta)>0$ such that\\ 1) For any $b>b(\eta)$, we have $\forall n\in[n(\eta),n_b]$,
	\begin{align}
		1-\eta\leq-\frac{\log\Pr\left[N_f(b)>n\right]}{\log n}\frac{\Lambda_1(\beta,\Pi)}{\lambda}\leq 1+\eta.\notag
	\end{align}
2) 
	\begin{align}
	\lim_{n\to\infty}\frac{\log\Pr\left[N_f(b)>n\right]}{n}=\log\left(\Pr\left[N_f>1|L_c=b\right]\right).\notag
	\end{align}
3) For any $b>b(\eta)$, we have $\forall t\in[n(\eta)b,n_bb]$,
	\begin{align}
		1-\eta\leq-\frac{\log\Pr\left[T_f(b)>t\right]}{\log t}\frac{\Lambda_1(\beta,\Pi)}{\lambda}\leq 1+\eta.\notag
	\end{align}
4)
	\begin{align}
	\lim_{n\to\infty}\frac{\log\Pr\left[T_f(b)>t\right]}{t}=\frac{1}{b}\log\left(\Pr\left[T_f>1|L_c=b\right]\right),\notag
	\end{align}
where
	\begin{align}
	n_b=\left(\Pr[N_f=1|L_c=b]\right)^{-1}.\label{nb}
	\end{align}
	The definition of $\Lambda_1(\beta,\Pi)$ can be found in Notation \ref{RateFunction}.
\end{theorem}
\begin{IEEEproof}
see Section \ref{fsenmproof}.
\end{IEEEproof}
\begin{remark}{\rm
From Equation (\ref{nb}) and by Lemma \ref{lemmanf}, we can obtain
\begin{align}
\lim_{b\to\infty}\frac{\log n_b}{b}=\Lambda_1(\beta,\Pi),\label{wa}
\end{align}
which implies that $n_b$ increases exponentially fast with the increase of maximum codeword length $b$. Since the waist of the heavy-tailed main body of the delay distribution is $n_bb$, we know that the waist also scales exponentially fast as we increase the maximum codeword length $b$. 
}\end{remark}
From Theorem \ref{fsenm} we know that even if the codeword length is bounded, the heavy-tailed main body could still play a dominant role. From Theorem \ref{ptp} we know that when $\lambda<\Lambda_1(\beta,\Pi)$ and $\beta>\gamma$, the throughput will vanish to zero as $b$ approaches infinity. Now we explore how fast the throughput vanishes to zero as $b$ increases. 

Let $\{L_{i}\}_{i\geq1}$ be the i.i.d. sequence of codeword lengths with distribution $L_c(b)$. Denote $T_i$ as the transmission delay of $L_i$. The throughput of this system is defined as $\Delta(b)=\lim_{n\to\infty}\sum_{i=1}^n\beta L_{i}/\sum_{i=1}^nT_i$.

\begin{theorem}[Throughput]\label{thput}In the case when decoder does not use memory and the codeword has variable length $L_c(b)$, if $\beta>\gamma$ and $\lambda<\Lambda_1(\beta,\Pi)$, we have
	\begin{align}
		-\limsup_{b\to\infty}\frac{\log\Delta(b)}{b}\geq\Lambda_1(\beta,\Pi)-\lambda.\notag
	\end{align}
	The definition of $\Lambda_1(\beta,\Pi)$ can be found in Notation \ref{RateFunction}.
\end{theorem}
\begin{IEEEproof}
see Section \ref{thputproof}.
\end{IEEEproof}
\begin{remark}{\rm
	Theorem \ref{thput} indicates that when code-rate $\beta$ is greater than channel capacity $\gamma$ and $\lambda<\Lambda_1(\beta,\Pi)$, as the maximum codeword length $b$ increases, the throughput vanishes to $0$ at least exponentially fast with rate $\Lambda_1(\beta,\Pi)-\lambda$.
}\end{remark}

\section{Simulations}\label{sims}
\begin{figure}[h]
\begin{center}
\includegraphics [width=3.5in]{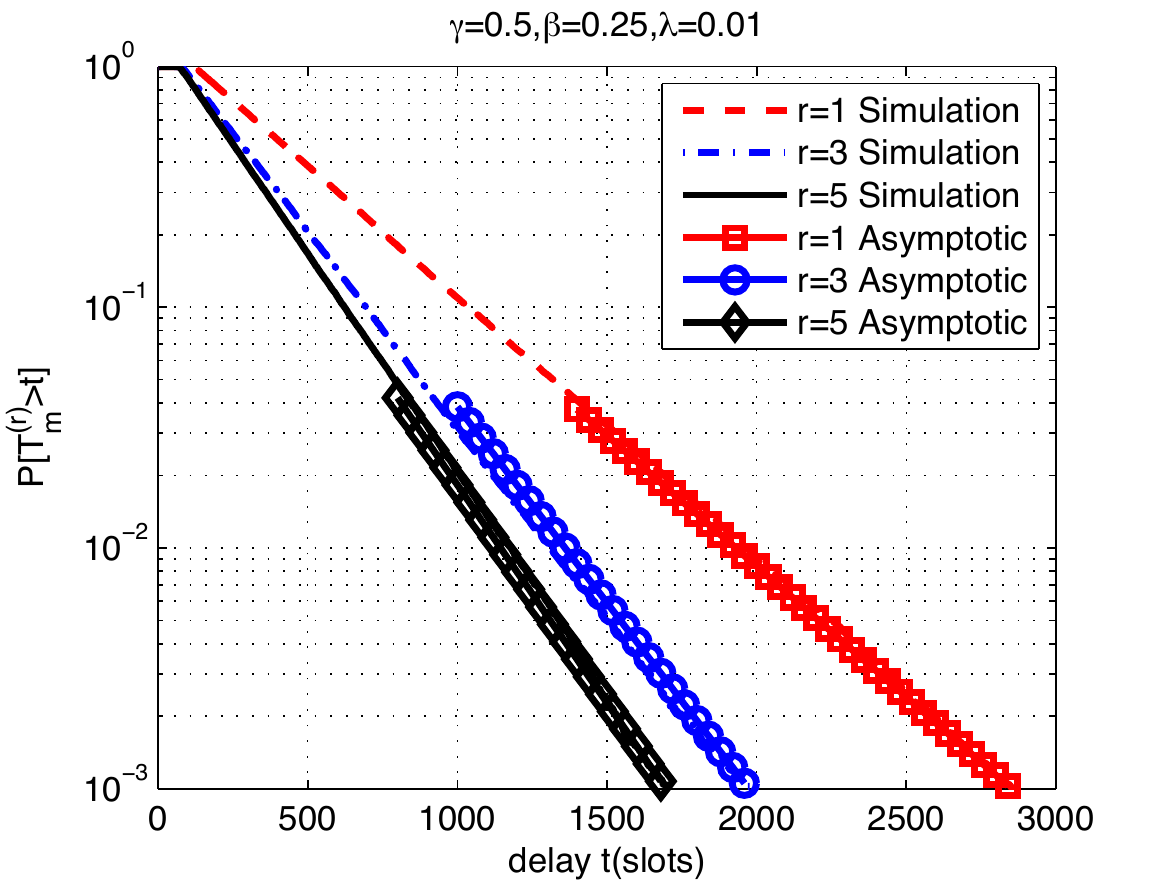}
\caption{Illustration for Example 1}
\label{umrcfig}
\end{center}
\end{figure}

In this section, we conduct simulations to verify our main results. As is evident from the following figures, the simulations match theoretical results well.
\begin{example}\label{example1}{\rm
In this example, we study the case when the decoder uses memory and the codeword length has infinite support. We assume that the channel is i.i.d.($k=0$). As shown in Theorem \ref{irc}, under the above assumptions, the delay distribution is always light-tailed. In order to verify this result, we assume that $L_c$ is geometrically distributed with mean $100$ ($\lambda=0.01$), and choose code-rate $\beta=0.5$ and channel capacity $\gamma=0.25$. By Theorem \ref{irc} we know that when $r=1$, the decay rate of delay is $\min\{\Lambda_1^o,\lambda\}=0.0025$; when $r=3$, the decay rate of delay is $\min\{\Lambda_1^o,\Lambda_3^o\}=\min\{\Lambda_2^o,\Lambda_3^o\}=0.0037$; when $r=5$, the decay rate of delay is $\min\{\Lambda_1^o,\Lambda_3^o\}=\min\{\Lambda_2^o,\Lambda_3^o\}=0.0042$. From Fig. \ref{umrcfig} we can see that the decay rate of delay increases when $r$ increases from $1$ to $5$, and the theoretical result is quite accurate.}
\end{example}

\begin{figure}[h]
\begin{center}
\includegraphics [width=3.5in]{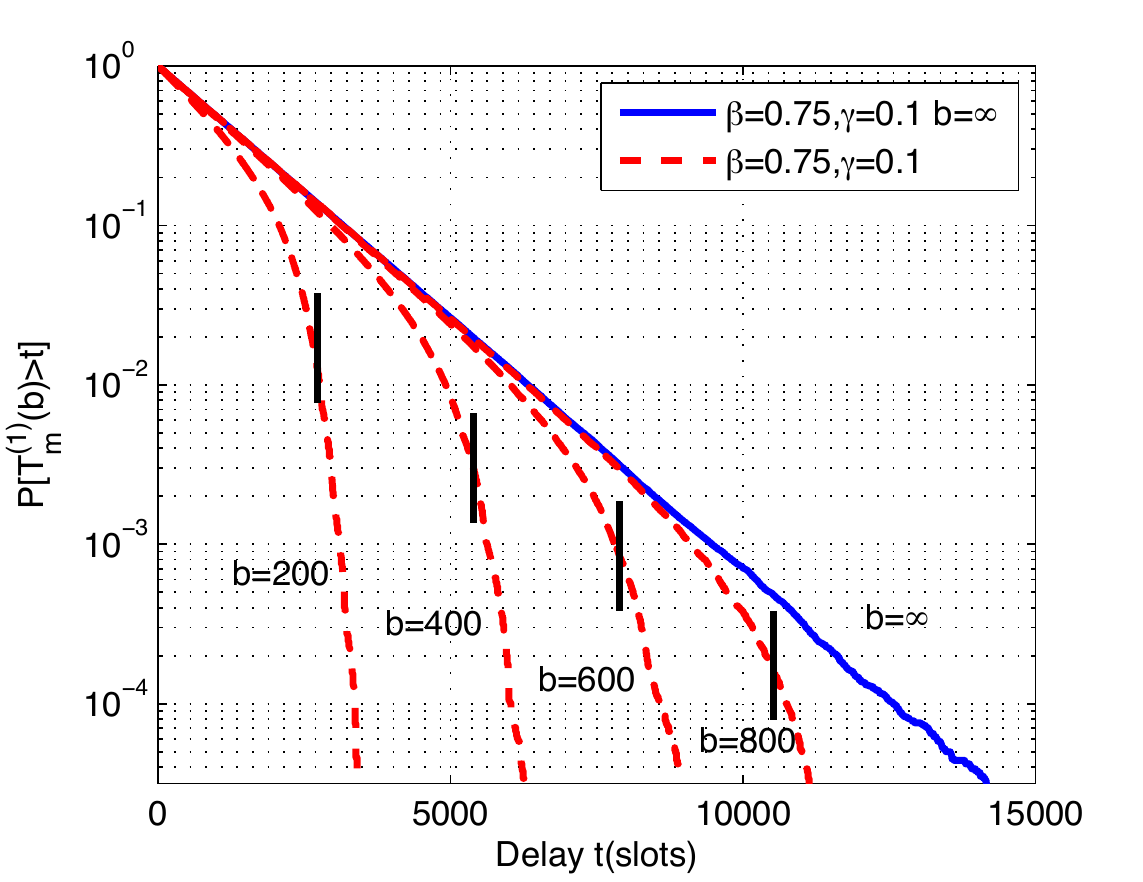}
\caption{Illustration for Example 2}
\label{umfsfig}
\end{center}
\end{figure}

\begin{example}\label{example2}{\rm
In this simulation, we study the case when the decoder uses memory and the codeword length has a finite support. We assume that the channel is i.i.d. ($k=0$), code-rate $\beta=0.75$, $\lambda=0.01$, $r=1$, and channel capacity $\gamma=0.1$. From these system parameters we can calculate $n_1^o=14$ and $\Lambda^b=\min\{\Lambda_1^o,\lambda\}=7.1429\times10^{-4}$. We choose four sets of maximum codeword length $b$ as $200,400,600,800$. Theorem \ref{fse} indicates that the delay distribution has a light-tailed main body with decay rate $\Lambda^b=7.1429\times10^{-4}$ and waist $n_bb=14\times b$. In Fig. \ref{umfsfig} we plot the delay distributions when $b=200,400,600,800$ together with the infinite support case when $b=\infty$, and we use a short solid line to indicate the waist of the light-tailed main body. As we can see from Fig. \ref{umfsfig}, the theoretical waists of the main bodies, which are $n_bb=14\times b=2800,5600,8400,11200$, are close to the simulation results.
}\end{example}

\begin{figure}[h]
\begin{center}
\includegraphics [width=3.5in]{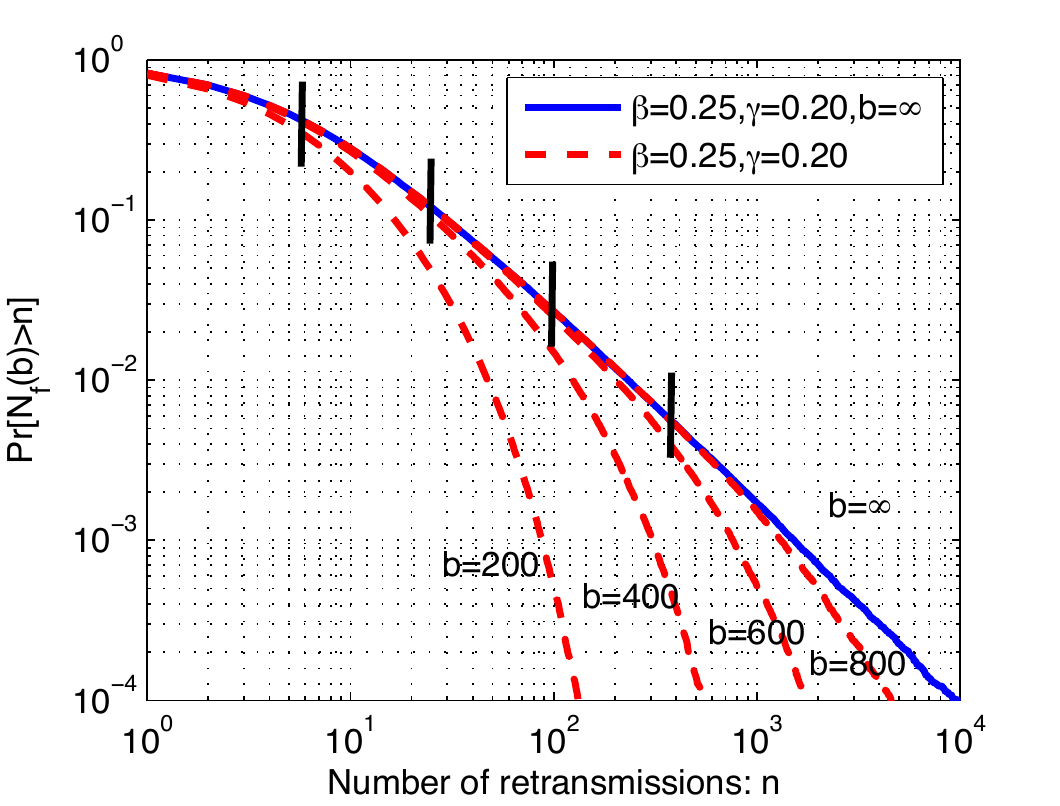}
\caption{Illustration for Example 3}
\label{dnumfsfig}
\end{center}
\end{figure}

\begin{example}\label{example3}{\rm
Now we use simulations to verify Theorem \ref{fsenm}. Theorem \ref{fsenm} says that when the decoder does not use memory, if code rate $\beta$ is greater than channel capacity $\gamma$ and the codeword length has a finite support, the distribution of delay as well as the distribution of number of retransmissions have a heavy-tailed main body and an exponential tail. The waist of the main body increases exponentially fast with the increase of maximum codeword length $b$. In this experiment, we set code-rate $\beta=0.25$, channel capacity $\gamma=0.20, k=0$, and $\lambda=0.01$. From these parameters we can get $\Lambda_1(\beta,\Pi)=0.0074$. We choose four sets of maximum codeword length $b$ as $200,400,600,800$. As Equation (\ref{wa}) indicates, the waist of the heavy-tailed main bodies of the number of retransmissions is $n_b\approx e^{b\Lambda_1(\beta,\Pi)}=4.3772,19.1595,83.8641,367.0865.$ In Fig. \ref{dnumfsfig}, we plot the distribution of the number of retransmissions when $b=200,400,600,800$ together with the infinite support case when $b=\infty$, and we use a short solid line to indicate the waist of the heavy-tailed main body. As can be seen from Fig. \ref{dnumfsfig}, the simulation matches with our theoretical result.
}\end{example}

\section{Proofs}
\subsection{Lemmas}
In order to prove the theorems, first we need the following three lemmas.
\begin{lemma}\label{Lemma1} 
\begin{align}
\Pr[N_m^{(1)}>n|L_c=l_c]=e^{-l_c\Lambda_n(\beta,\Pi){\bf1}\left(n\geq\alpha\right)+g_{n}(l_c)},\notag
\end{align}
where
\begin{align}
g_n(l_c)\in\left\{\begin{array}{ll}
o(l_c)&\text{ if } n\geq\alpha\\
o(1)  &\text{ otherwise}
\end{array}\right..\notag
\end{align}
\end{lemma}
\begin{IEEEproof}
First we consider the case when $k\geq1$. By Definition \ref{Nm}, we have
	\begin{align}
		 \Pr{\big [}&N_m^{(1)}>n|L_c=l_c{\big ]}\notag\\
		=\Pr{\Bigg [}&\sum_{i=1}^{L_c}{\bf 1}{\Bigg (}\sum_{j=1}^nX_{(j-1)L_c+i}\geq1{\Bigg )}\leq\beta L_c{\Bigg |}L_c{\Bigg ]}\notag\\
		=\Pr{\Bigg [}&\sum_{i=1}^{L_c}{\bf 1}{\Bigg (}\sum_{j=1}^nX_{(j-1)L_c+i}=0{\Bigg )}>(1-\beta)L_c{\Bigg |}L_c{\Bigg ]}\notag\\
		=\mathbb{E}{\Bigg [}&\Pr{\Bigg [}\sum_{i=1}^{L_c}\prod_{j=1}^n{\bf 1}{\bigg (}X_{(j-1)L_c+i}=0{\bigg )}\notag\\
		&\text{ }\text{ }\text{ }\text{ }\text{ }\text{ }>(1-\beta)L_c{\Bigg |}L_c,\bigcup_{j=1}^n\mathcal{E}_j{\Bigg ]}{\Bigg |}L_c{\Bigg ]},\label{nm1}
	\end{align}
	where $\mathcal{E}_j\triangleq\left\{X_{(j-1)L_c+1},\ldots,X_{(j-1)L_c+k}\right\}$, $1\leq j\leq n$.\\
	Let $Y_{in}=\left[Y_i,Y_{L_c+i},\ldots,Y_{(n-1)L_c+i}\right]$ and
	\begin{align}
		 f_n\left(Y_{in}\right)=\prod_{j=1}^nf\left(Y_{(j-1)L_c+i}\right)\notag.
	\end{align}
	If $L_c>k$, then given $\bigcup_{j=1}^n\mathcal{E}_j$, $\left\{Y_{in}\right\}_{k<i\leq L_c}$ forms a Markov chain with state space $\{\{0,1\}^k\}^n$ and probability transition matrix $\Pi^{\otimes n}$. We further observe that if $L_c>k$, we have the following relationship
	\begin{align}
	&{\Bigg\{}\sum_{i=1+k}^{L_c}\prod_{j=1}^n{\bf 1}{\bigg(}X_{(j-1)L_c+i}=0{\bigg)}>(1-\beta)L_c{\Bigg\}}\notag\\
	&\subseteq
	{\Bigg\{}\sum_{i=1}^{L_c}\prod_{j=1}^n{\bf 1}{\bigg(}X_{(j-1)L_c+i}=0{\bigg)}>(1-\beta)L_c{\Bigg\}}\notag\\
	&\subseteq
	{\Bigg\{}\sum_{i=1+k}^{L_c}\prod_{j=1}^n{\bf 1}{\bigg(}X_{(j-1)L_c+i}=0{\bigg)}>(1-\beta)L_c-k{\Bigg\}}\notag.
	\end{align}
	Using the above observation, we can construct upper and lower bounds as follows.
	\begin{align}
	&\Pr{\Bigg [}\sum_{i=1}^{L_c}\prod_{j=1}^n{\bf 1}{\Bigg(}X_{(j-1)L_c+i}=0{\Bigg)}>(1-\beta)L_c{\Bigg|}L_c,\bigcup_{j=1}^n\mathcal{E}_j{\Bigg ]}\notag\\
	&\geq\Pr{\Bigg [}\sum_{i=1+k}^{L_c}f_n(Y_{in})>(1-\beta)L_c{\Bigg|}L_c,\bigcup_{j=1}^n\mathcal{E}_j{\Bigg ]},\notag\\
	&\Pr{\Bigg [}\sum_{i=1}^{L_c}\prod_{j=1}^n{\bf 1}{\Bigg(}X_{(j-1)L_c+i}=0{\Bigg)}>(1-\beta)L_c{\Bigg|}L_c,\bigcup_{j=1}^n\mathcal{E}_j{\Bigg ]}\notag\\
	&\leq\Pr{\Bigg [}\sum_{i=1+k}^{L_c}f_n(Y_{in})>(1-\beta)L_c-k{\Bigg|}L_c,\bigcup_{j=1}^n\mathcal{E}_j{\Bigg ]}.\label{late1}
	\end{align}
	By a direct application of Theorem 3.1.2 in \cite{Dembozeitouni}, we know that for a given $\varepsilon>0$ and any values of $\bigcup_{j=1}^n\mathcal{E}_j$, we can find $l_\varepsilon$ such that 	 
	\begin{align}
	&\Pr{\Bigg [}\sum_{i=1+k}^{L_c}f_n(Y_{in})>(1-\beta)L_c{\Bigg|}L_c=l_c{\Bigg ]}\notag\\
	&\geq e^{-\inf_{1-\omega>1-\beta}\Lambda_n(\beta,\Pi)(1+\varepsilon)l_c},\label{Mlower1}\\
	&\Pr{\Bigg [}\sum_{i=1+k}^{L_c}f_n(Y_{in})>(1-\beta)L_c-k{\Bigg|}L_c=l_c{\Bigg ]}\notag\\
	&\leq e^{-\inf_{1-\omega>1-\beta}\Lambda_n(\beta,\Pi)(1-\varepsilon)l_c},\label{Mupper1}
	\end{align}
	whenever $l_c>l_\varepsilon$. 
	Since $\Lambda_n(\omega,\Pi)$ is a large deviation rate function, from \cite{Dembozeitouni} we know that
	\begin{align}
		\inf_{1-\omega>1-\beta}\Lambda_n(\omega,\Pi)&=\left\{
			\begin{array}{ll}
				\Lambda_n(\beta,\Pi)&\text{ if }\mu_n<\beta\\
				0&\text{ otherwise }
			\end{array}
		\right.\notag\\
		&=\Lambda_n(\beta,\Pi){\bf 1}(n\geq\alpha).\label{fojing}
	\end{align}
	The upper and lower bounds (\ref{Mupper1}) and (\ref{Mlower1}), together with Equation (\ref{nm1}), (\ref{late1}) and (\ref{fojing}), imply that
	\begin{align}
	 -\liminf_{l_c\to\infty}&\frac{\log\Pr\left[N_m>n|L_c=l_c\right]}{l_c}\notag\\
	 &\leq(1+\varepsilon)\Lambda_n(\beta,\Pi){\bf 1}(n\geq\alpha),\notag\\
	 -\limsup_{l_c\to\infty}&\frac{\log\Pr\left[N_m>n|L_c=l_c\right]}{l_c}\notag\\
	 &\geq(1-\varepsilon)\Lambda_n(\beta,\Pi){\bf 1}(n\geq\alpha),\notag
	\end{align}
	which, with $\varepsilon\to0$, completes the proof when $k\geq1$. Next, let us consider the case when $k=0$. 
	
	In this memoryless channel case, for a single bit in the codeword, after $n^{\text{th}}$ transmission, the probability that this bit is successfully received is $1-(1-\gamma)^n$. 
	Therefore equivalently, we can consider a single transmission in a memoryless channel with erasure probability $(1-\gamma)^n$. Then, by a direct application of G\"{a}rtner-Ellis theorem (Theorem 2.3.6 in \cite{Dembozeitouni}), we have, for any $n\geq1$,
	\begin{align}
	\lim_{l_c\to\infty}\frac{\log\Pr[N_m>n|L_c=l_c]}{l_c}=-\Lambda_n(\beta,\Pi){\bf1}\left(n\geq\alpha\right),\notag
	\end{align}
	where
	\begin{align}
	&\Lambda_n(\beta,\Pi)=\sup_{\theta}\left\{\theta(1-\beta)-\log\left(\mathbb{E}\left[X_ie^{\theta X_i}\right]\right)\right\}\notag\\
	&=\sup_{\theta}\{\theta(1-\beta)-\log\left(1-(1-\gamma)^n+(1-\gamma)^ne^\theta\right)\}\notag\\
	&=\beta\log\frac{\beta}{1-(1-\gamma)^n}+(1-\beta)\log\frac{1-\beta}{(1-\gamma)^n},\notag
	\end{align}
	and $\alpha=\left\lceil\frac{\log(1-\beta)}{\log(1-\gamma)}\right\rceil$.
\end{IEEEproof}

\begin{lemma}\label{pmt}
Assume $b$ is a function of $t$, which satisfies $b\triangleq b(t)>\frac{t}{y}$. Then, for any $x,y\in\mathbb{N}$ we have
	\begin{align}
	&\lim_{t\to\infty}\frac{\log\Pr\left[N_m^{(1)}>x,\frac{t}{y+1}<L_c(b)<\frac{t}{y}\right]}{t}\notag\\
	&=-\frac{\lambda+\Lambda_x(\beta,\Pi){\bf1}(x\geq\alpha)}{y+1}.\notag
	\end{align}
\end{lemma}
\begin{IEEEproof}
\begin{align}
	&\Pr\left[N_m^{(1)}>x,\frac{t}{y+1}<L_c(b)<\frac{t}{y}\right]\notag\\
	=&\sum_{l_c=\lceil t/(y+1)\rceil}^{\lfloor t/y\rfloor}\Pr\left[N_m^{(1)}>x{\bigg|}L_c(b)=l_c\right]\Pr\left[L_c(b)=l_c\right]\notag\\
	\leq&\Pr\left[N_m^{(1)}>x{\bigg|}L_c(b)=\left\lceil\frac{t}{y+1}\right\rceil\right]\Pr\left[\frac{t}{y+1}<L_c(b)<\frac{t}{y}\right]\label{pmt1}.
\end{align}
From Lemma \ref{Lemma1} we know that 
\begin{align}
	&\Pr\left[N_m^{(1)}>x{\bigg|}L_c(b)=\left\lceil\frac{t}{y+1}\right\rceil\right]\notag\\
	=&e^{-\left\lceil\frac{t}{y+1}\right\rceil\Lambda_x(\beta,\Pi){\bf1}\left(x\geq\alpha\right)+g_{x}(l_c)}.\label{pmt2}
\end{align}
Since $b=b(t)>\frac{t}{y}$, by the definition of $L_c(b)$, we can easily obtain
\begin{align}
	\lim_{t\to\infty}\frac{\log\Pr\left[\frac{t}{y+1}<L_c(b)<\frac{t}{y}\right]}{t}=-\frac{\lambda}{y+1}.\label{pmt3}
\end{align}
Combining Equation (\ref{pmt3}), (\ref{pmt2}) and (\ref{pmt1}), we get 
\begin{align}
&\limsup_{t\to\infty}\frac{\log\Pr\left[N_m^{(1)}>x,\frac{t}{y+1}<L_c(b)<\frac{t}{y}\right]}{t}\notag\\
\leq&-(\lambda+\Lambda_x(\beta,\Pi){\bf1}(x\geq\alpha))/(y+1).\notag
\end{align}
The lower bound can be constructed in a similar manner.
\end{IEEEproof}

\begin{lemma}\label{lemmanf}
\begin{enumerate}
\item if $\beta>\gamma$, then
	\begin{align}
		&\Pr\left[N_f>n|L_c=l_c\right]=\left(1-e^{-l_c\Lambda_1(\beta,\Pi)(1+g(l_c))}\right)^n,\notag
	\end{align}
where $g(l_c)\in o(1)$ as $l_c\to\infty$.
\item if $\beta<\gamma$, then
	\begin{align}
		&\Pr\left[N_f>n|L_c=l_c\right]=e^{-nl_c\Lambda_1(\beta,\Pi)(1+s(l_c))},\notag
	\end{align}
where $s(l_c)\in o(1)$ as $l_c\to\infty$.
\end{enumerate}
\end{lemma}
\begin{IEEEproof}
From Definition \ref{Nf} we know
\begin{align}
&\Pr[N_f >n | L_c]\notag\\
=& \Pr \left[ \bigcap_{1\leq j \leq n} \left\{
\sum_{i=(j-1)L_c+1}^{jL_c}X_i \leq \beta L_c \right\} {\Bigg |} L_c
\right].\notag\\
=&\mathbb{E}\left[\Pr \left[ \bigcap_{1\leq j \leq n} \left\{
\sum_{i=(j-1)L_c+1}^{jL_c}X_i \leq \beta L_c \right\} {\Bigg |}\bigcup_{j=1}^n\mathcal{E}_j,L_c
\right]{\Bigg|}L_c\right]\notag\\
=&\mathbb{E}\left[\prod_{j=1}^n\Pr \left[
\sum_{i=(j-1)L_c+1}^{jL_c}X_i \leq \beta L_c  {\Bigg |}\bigcup_{i=1}^n\mathcal{E}_i,L_c
\right]{\Bigg|}L_c\right],\label{nfeq1}
\end{align}
where $\mathcal{E}_j\triangleq\left\{X_{(j-1)L_c+1},\ldots,X_{(j-1)L_c+k}\right\}$, $1\leq j\leq n$.
The last equation is due to the Markov property of the channel states. Observe that if $L_c>k$, for any $1\leq j\leq n$,
\begin{align}
	&\left\{\sum_{i=(j-1)L_c+1+k}^{jL_c}X_i \leq \beta L_c\right\}\notag\\
	\subseteq&
	\left\{\sum_{i=(j-1)L_c+1}^{jL_c}X_i \leq \beta L_c\right\}\notag\\
	\subseteq&
	\left\{\sum_{i=(j-1)L_c+1+k}^{jL_c}X_i \leq \beta L_c-k\right\}\notag,
\end{align}
which further yields
\begin{align}
&\Pr \left[
\sum_{i=(j-1)L_c+1+k}^{jL_c}X_i \leq \beta L_c -k {\Bigg |}\bigcup_{i=1}^n\mathcal{E}_i,L_c
\right]\notag\\
\geq&\Pr \left[
\sum_{i=(j-1)L_c+1}^{jL_c}X_i \leq \beta L_c  {\Bigg |}\bigcup_{i=1}^n\mathcal{E}_i,L_c
\right]\notag\\
\geq&\Pr \left[
\sum_{i=(j-1)L_c+1+k}^{jL_c}X_i \leq \beta L_c  {\Bigg |}\bigcup_{i=1}^n\mathcal{E}_i,L_c
\right].\notag
\end{align}
Similarly as the proof of Lemma \ref{Lemma1}, by Theorem 3.1.2 in \cite{Dembozeitouni}, we obtain, for any $1\leq j\leq n$
\begin{align}
&\lim_{l_c\to\infty}\frac{\log\Pr \left[
\sum_{i=(j-1)L_c+1}^{jL_c}X_i > \beta L_c  {\big|}\bigcup_{i=1}^n\mathcal{E}_i,L_c=l_c
\right]}{l_c}\notag\\
&=-\Lambda_1(\beta,\Pi){\bf 1}(\beta>\gamma),\notag\\
&\lim_{l_c\to\infty}\frac{\log\Pr \left[
\sum_{i=(j-1)L_c+1}^{jL_c}X_i \leq \beta L_c  {\big|}\bigcup_{i=1}^n\mathcal{E}_i,L_c=l_c
\right]}{l_c}\notag\\
&=-\Lambda_1(\beta,\Pi){\bf 1}(\beta<\gamma),\notag
\end{align}
which, by combining Equation \ref{nfeq1}, completes the proof.
\end{IEEEproof}

\subsection{Proof of Theorem \ref{irc}}\label{ircproof}
\begin{IEEEproof}[proof of  Theorem \ref{irc}]
Observe that
\begin{align}
 &\Pr\left[T_m^{(r)}>t\right]\notag\\
=&\sum_{h=r}^\infty\Pr\left[T_m^{(r)}>t,\frac{tr}{h+1}<L_c\leq\frac{tr}{h}\right]+\Pr\left[T_m^{(r)}>t,L_c>t\right]\notag\\
=&\sum_{n=1}^\infty\sum_{h=nr}^{(n+1)r-1}\Pr\left[N_m^{(r)}>h,\frac{tr}{h+1}<L_c\leq\frac{tr}{h}\right]\notag\\
 &+\Pr\left[T_m^{(r)}>t,L_c>t\right].\label{irceq}
\end{align}
Let us first focus on the first part of Equation (\ref{irceq}). Denote $\Pr_{ntr}=\sum_{h=nr}^{(n+1)r-1}\Pr\left[N_m^{(r)}>h,\frac{tr}{h+1}<L_c\leq\frac{tr}{h}\right]$, then it is easy to check that
\begin{align}
&\Pr_{ntr}\leq\Pr\left[N_m^{(1)}>n,\frac{t}{n+1}<L_c\leq\frac{t}{n}\right],\notag\\
&\Pr_{ntr}\geq\Pr\left[N_m^{(1)}>n+1,\frac{t}{n+1}<L_c\leq\frac{t}{n}\right],\notag
\end{align}
which, by Lemma \ref{pmt}, yield
\begin{align}
\limsup_{t\to\infty}\frac{\log\Pr_{ntr}}{t}&\leq-\frac{\lambda+\Lambda_n(\beta,\Pi){\bf 1}(n\geq\alpha)}{n+1},\notag\\
\liminf_{t\to\infty}\frac{\log\Pr_{ntr}}{t}&\geq-\frac{\lambda+\Lambda_{n+1}(\beta,\Pi){\bf 1}(n\geq\alpha-1)}{n+1}.\label{2bounds}
\end{align}
For the second part of Equation (\ref{irceq}), we have, by the definition of $L_c$,
\begin{align}
&-\lim_{t\to\infty}\frac{\log\Pr\left[T_m^{(r)}>t,L_c>t\right]}{t}\notag\\
=&\left\{\begin{array}{ll}
		\lim_{t\to\infty}\frac{\log\Pr[L_c>t]}{t}&\text{ if }\beta>\gamma\\
		\lim_{t\to\infty}\frac{\log\Pr[L_c>tr/\left\lceil\frac{r\beta}{\gamma}\right\rceil]}{t}&\text{ if }\beta\leq\gamma
		\end{array}\right.\notag\\
=&\left\{
	\begin{array}{ll}
	\lambda&\text{ if }\beta>\gamma\\
	\frac{\lambda r}{\lceil r\beta/\gamma\rceil}&\text{ if }\beta\leq\gamma
	\end{array}
\right.=\Lambda_3^o.\label{tmrl}
\end{align}
Combining Equation (\ref{irceq}), (\ref{2bounds}) and (\ref{tmrl}), we get
\begin{align}
&\limsup_{t\to\infty}\frac{\log\Pr\left[T_m^{(r)}>t\right]}{t}\notag\\
=&\max{\Big\{}\limsup_{t\to\infty}\frac{\log\sum_{n=1}^\infty\Pr_{ntr}}{t},\notag\\
&\text{ }\text{ }\text{ }\text{ }\text{ }\lim_{t\to\infty}\frac{\log\Pr\left[T_m^{(r)}>t,L_c>t\right]}{t}{\Bigg\}}\notag\\
\overset{(a)}{\leq}&\max\left\{-\inf_{n}\frac{\lambda+\Lambda_n(\beta,\Pi){\bf 1}(n\geq\alpha)}{n+1},-\Lambda_3^o\right\}\notag\\
=&-\min\{\Lambda_1^o,\Lambda_3^o\}.\label{trick}
\end{align}
The lower bound can be found in a similar manner. Notice that inequality (a) in the preceding equation is true because $\Pr_{ntr}$ is nonzero only for a finite number of $n$, which is due to the fact that $L_c$ cannot be less than $1$. 

In the special case when $r=1$, by Lemma \ref{pmt} and the definition of $L_c$, we have
\begin{align}
&\lim_{t\to\infty}\frac{\log\Pr_{ntr}}{t}=-\frac{\lambda+\Lambda_n(\beta,\Pi){\bf 1}(n\geq\alpha)}{n+1},\notag\\
&\lim_{t\to\infty}\frac{\log\Pr\left[T_m^{(1)}>t,L_c>t\right]}{t}=-\lambda,
\end{align}
which, by combining Equation (\ref{irceq}), completes the proof.
\end{IEEEproof}

\subsection{Proof of Theorem \ref{fse}}\label{fseproof}
\begin{IEEEproof}[proof of Theorem \ref{fse}]
From the definition of $n^o_2$, $\Lambda_1^o$ and $\Lambda_2^o$ in Notation \ref{lambdas} and by Lemma \ref{pmt}, we can obtain, for any $b_0>0$,
\begin{align}
\limsup_{t\to\infty}-\frac{\log\Pr\left[T_m^{(r)}>t,\frac{t}{n^o_2+1}<L_c\left(\frac{t}{n^o_2}+b_0\right)\leq\frac{t}{n^o_2}\right]}{t}\leq\Lambda_2^o,\notag\\
\liminf_{t\to\infty}-\frac{\log\Pr\left[T_m^{(r)}>t,\frac{t}{n^o_2+1}<L_c\left(\frac{t}{n^o_2}+b_0\right)\leq\frac{t}{n^o_2}\right]}{t}\geq\Lambda_1^o.\notag
\end{align}
Then, for any $\eta>0$ and for any $b_0>0$, we can find $t(\eta)$ such that 
\begin{align}
-\frac{\log\Pr\left[T_m^{(r)}>t,\frac{t}{n^o_2+1}<L_c\left(\frac{t}{n^o_2}+b_0\right)\leq\frac{t}{n^o_2}\right]}{t}\leq(1+\eta)\Lambda_2^o,\notag\\
-\frac{\log\Pr\left[T_m^{(r)}>t,\frac{t}{n^o_2+1}<L_c\left(\frac{t}{n^o_2}+b_0\right)\leq\frac{t}{n^o_2}\right]}{t}\geq(1-\eta)\Lambda_1^o,\notag
\end{align}
whenever $t>t(\eta)$. We denote $b(\eta)\triangleq\frac{t(\eta)}{n^o_2}+b_0$. In other words, for any $b>b(\eta)$, whenever $t\in[(b-b_0)n^o_2,bn^o_2]$,
\begin{align}
-\frac{\log\Pr\left[T_m^{(r)}>t,\frac{t}{n^o_2+1}<L_c\left(b\right)\leq\frac{t}{n^o_2}\right]}{t}\leq(1+\eta)\Lambda_2^o,\notag\\
-\frac{\log\Pr\left[T_m^{(r)}>t,\frac{t}{n^o_2+1}<L_c\left(b\right)\leq\frac{t}{n^o_2}\right]}{t}\geq(1-\eta)\Lambda_1^o,\notag
\end{align}
which, by using the same technique as in Equation (\ref{trick}), completes the proof of the first part. The second part of Theorem \ref{fse} follows by noting that
\begin{align}
\lim_{t\to\infty}-\frac{\log\Pr\left[T_m^{(1)}>t,\frac{t}{n^o_1+1}<L_c\left(\frac{t}{n^o_1}+b_0\right)\leq\frac{t}{n^o_1}\right]}{t}=\Lambda^o_1,\notag
\end{align}
where the definition of $n_1^o$ can be found in Notation \ref{lambdas}.

\end{IEEEproof}

\subsection{Proof of Theorem \ref{ptp}}\label{ptpproof}
\begin{IEEEproof}[proof of Theorem \ref{ptp}]
1) If $\beta>\gamma$, by Lemma \ref{lemmanf}, for any $\varepsilon>0$, we can find $l_{\varepsilon}$ such that
\begin{align}
\Pr[N_f>n|L_c=l_c]\geq\left(1-e^{-l_c\Lambda_1(\beta,\Pi)(1-\varepsilon)}\right)^n,\notag
\end{align}
whenever $l_c>l_\varepsilon$. Then we have, for $n$ large enough,
\begin{align}
\Pr [N_f & >n]= \mathbb{E}\left[\Pr [N_f >n | L_c]\right]
\notag\\
& \geq \mathbb{E}\left[ L_c>l_{\epsilon} , \left(1- e^{-
\Lambda_1(\beta,\Pi)(1- \epsilon)L_c} \right)^n \right]\notag\\
&\geq \mathbb{E}{\Big[}\frac{\log n}{\Lambda_1(\beta,\Pi)(1-\epsilon)} <
L_c < \frac{\log n}{\Lambda_1(\beta,\Pi)(1-\epsilon)}+z , \notag\\
&\quad \quad \; \left(1- e^{-
\Lambda_1(\beta,\Pi)(1- \epsilon)L_c} \right)^n {\Big]}\notag\\
&\geq \mathbb{E}{\Big[}\frac{\log n }{\Lambda_1(\beta,\Pi)(1-\epsilon)}<
L_c < \frac{\log n }{\Lambda_1(\beta,\Pi)(1-\epsilon)}+z , \notag\\
&\quad \quad \; \left(1- e^{- (\log n )} \right)^n {\Big]} \notag\\
&\geq e^{-\lambda(1+\epsilon)\frac{\log
n}{\Lambda_1(\beta,\Pi)(1-\epsilon)}} \left(1- e^{- (\log n )} \right)^n.
\notag
\end{align}
Taking logarithms on both sides of the preceding inequality, we get
\begin{align}
   \liminf_{n \to \infty}\frac{\log\Pr[N_f>n]}{\log n} \geq
   -\frac{\lambda(1+\epsilon)}{\Lambda_1(\beta,\Pi)(1-\epsilon)},\notag
\end{align}
which, when $\epsilon \to 0$, results in the lower bound. 

Next, we prove the upper bound. Using the same technique as in the proof of the lower bound, and by the definition of $L_c$, we can find $l_\varepsilon$ such that

\begin{align}
\Pr [N_f &>n] \leq  \Pr\left[ L_c>l_{\epsilon}, \left(1- e^{-
\Lambda_1(\beta,\Pi)(1+\epsilon)L_c} \right)^n \right] \notag\\
& \quad \;+ \Pr[N_f>n, L_c \leq l_{\epsilon}] \notag\\
&\leq \sum_{l=l_{\epsilon}}^{\infty}\left(1- e^{-
\Lambda_1(\beta,\Pi)(1+\epsilon)l} \right)^n  \Pr[L_c= l] + O(e^{- \xi
n}), \notag\\
&\leq O\left(\int_{0}^{\infty}\left(1- e^{-
\Lambda_1(\beta,\Pi)(1+\epsilon)x} \right)^n e^{-\lambda(1-\epsilon)
x}dx\right) \notag\\
&\quad  + O(e^{- \xi n}). \notag
\end{align}
Computing the integrated in the preceding inequality, we obtain
\begin{align}
   \limsup_{n \to \infty}\frac{\log\Pr[N_f>n]}{\log n} \leq
   -\frac{\lambda(1-\epsilon)}{\Lambda_1(\beta,\Pi)(1+\epsilon)},\notag
\end{align}
which, with $\epsilon \to 0$, proves the upper bound.

Now, we prove the result for $\Pr[T_f>t]$. The upper bound follows
by noting that
\begin{align}
 \Pr[T_f>t] & \leq \Pr[N_fL_c>t, L_c\leq h \log t] + \Pr[L_c>h\log
 t] \notag\\
 &\leq \Pr[N_f>t/(h \log t)] + \Pr[L_c>h\log t], \notag
\end{align}
where $\lim_{t\to \infty}\log\Pr[N_f>t/(h \log t)]/\log t =
\lambda/\Lambda_1(\beta,\Pi)$, and $\Pr[L_c>h\log t]=o\left(\Pr[N_f>t/(h
\log t)] \right)$ for $h$ large enough.

The lower bound follows by noting that, for some $l_2>l_1>0$ with
$\Pr[l_1<L_c<l_2]>0$,
\begin{align}
 \Pr[T_f>t] & \geq \Pr[N_fL_c>t, l_1<L_c <l_2]  \notag\\
 &\geq \Pr[N_f>t/l_1]\Pr[ l_1<L_c <l_2]. \notag
\end{align}
2) Observe that
\begin{align}
	&\Pr[T_f>t]\notag\\
	=&\Pr[L_c>t]+\sum_{n=1}^\infty\Pr\left[N_f>n,\frac{t}{n+1}<L_c\leq\frac{t}{n}\right].\label{ptp2eq}
\end{align}
Using the same technique as in the proof of Lemma \ref{pmt} and by Lemma \ref{lemmanf}, we have, when $\beta<\gamma$,
\begin{align}
&\lim_{t\to\infty}\frac{\log\Pr\left[N_f>n,\frac{t}{n+1}<L_c\leq\frac{t}{n}\right]}{t}=-\frac{n\Lambda_1(\beta,\Pi)+\lambda}{n+1},\notag\\
&\lim_{t\to\infty}\frac{\log\Pr[L_c>t]}{t}=-\lambda,\notag
\end{align}
which, by combining Equation \ref{ptp2eq} and using the same technique as in Equation (\ref{trick}), yield
\begin{align}
\lim_{t\to\infty}\frac{\log\Pr[T_f>t]}{t}&=-\min\left\{\inf_{n\in\mathbb{N}}\left\{\frac{n\Lambda_1(\beta,\Pi)+\lambda}{n+1}\right\},\lambda\right\}\notag\\
&=-\min\{\Lambda_1(\beta,\Pi),\lambda\}.\notag
\end{align}
\end{IEEEproof}

\subsection{Proof of Theorem \ref{fsenm}}\label{fsenmproof}
\begin{IEEEproof}[proof of Theorem \ref{fsenm}]
1) By Lemma \ref{lemmanf} we know that when $\beta>\gamma$
	\begin{align}
		&\Pr\left[N_f>n|L_c=l_c\right]=\left(1-e^{-l_c\Lambda_1(\beta,\Pi)(1+g(l_c))}\right)^n,\notag
	\end{align}
	with $g(l_c)\in o(1)$ as $l_c\to\infty$. Let us denote $l_n$ as the root of the function $l_c(1+g(l_c))-\frac{\log n}{\Lambda_1(\beta,\Pi)}$. In other words,
	\begin{align}
	l_n(1+g(l_n))=\frac{\log n}{\Lambda_1(\beta,\Pi)}.\notag
	\end{align}
	For any $b_0>0$, we have
	\begin{align}
	&\Pr\left[N_f>n,l_n-z<L_c(l_n+b_0)\leq l_n\right]\notag\\
	\geq&\Pr\left[N_f>n|L_c=l_n\right]\Pr\left[l_n-z<L_c(l_n+b_0)\leq l_n\right].\label{51}
	\end{align}
	Note that, by Lemma \ref{lemmanf},
	\begin{align}
	&\lim_{n\to\infty}\frac{\log\Pr\left[N_f>n|L_c=l_n\right]}{\log n}\notag\\
	=&\lim_{n\to\infty}\frac{\log\left(1-e^{-l_n\Lambda_1(\beta,\Pi)(1+g(l_c))}\right)^n}{\log n}\notag\\
	=&\lim_{n\to\infty}\frac{n\log(1+\frac{1}{n})}{\log n}=0.\label{52}
	\end{align}
	Also, by the definition of $L_c(b)$,
	\begin{align}
	&\lim_{n\to\infty}\frac{\log \Pr\left[l_n-z<L_c(l_n+b_0)\leq l_n\right]}{\log n}\notag\\
	=&\lim_{n\to\infty}\frac{\log \Pr\left[l_n-z<L_c(l_n+b_0)\leq l_n\right]}{l_n}\frac{l_n}{\log n}\notag\\
	=&-\lambda\frac{1}{\Lambda_1(\beta,\Pi)}.\label{53}
	\end{align}
	Combining Equation (\ref{51}), (\ref{52}) and (\ref{53}), we get
	\begin{align}
	&\lim_{n\to\infty}\frac{\log\Pr\left[N_f>n,l_n-z<L_c(l_n+b_0)\leq l_n\right]}{\log n}\notag\\
	=&-\frac{\lambda}{\Lambda_1(\beta,\Pi)}.\notag
	\end{align}
	Therefore, for any $\eta>0$, we can find a $n_1(\eta)$ such that
	\begin{align}
	&\lim_{n\to\infty}\frac{\log\Pr\left[N_f>n,l_n-z<L_c(l_n+b_0)\leq l_n\right]}{\log n}\notag\\
	\geq&-\frac{\lambda}{\Lambda_1(\beta,\Pi)}(1+\eta),\label{hte1}
	\end{align}
	whenever $n>n_1(\eta)$. Also, by Theorem \ref{ptp}, we can find $n_2(\eta)$ such that
	\begin{align}
	\lim_{n\to\infty}\frac{\log\Pr\left[N_f>n\right]}{\log n}\leq-\frac{\lambda}{\Lambda_1(\beta,\Pi)}(1-\eta).\label{hte2}
	\end{align}
	Let $n(\eta)\triangleq\max\{n_1(\alpha),n_2(\alpha)\}$ and $b(\eta)\triangleq l_{n(\eta)}+b_0$. By combining Equation (\ref{hte1}) and (\ref{hte2}), we know that for any $\eta>0$, we can find $b(\eta)$ such that for any $b>b(\eta)$,
	\begin{align}
	\limsup_{n\to\infty}\frac{\log\Pr\left[N_f(b)>n\right]}{\log n}&\leq\lim_{n\to\infty}\frac{\log\Pr\left[N_f>n\right]}{\log n}\notag\\
	&\leq-\frac{\lambda}{\Lambda_1(\beta,\Pi)}(1-\eta),\notag
	\end{align}
	and
	\begin{align}
	&\liminf_{n\to\infty}\frac{\log\Pr\left[N_f(b)>n\right]}{\log n}\notag\\
	\geq&	\lim_{n\to\infty}\frac{\log\Pr\left[N_f>n,l_n-z<L_c(l_n+b_0)\leq l_n\right]}{\log n}\notag\\
	\geq&-\frac{\lambda}{\Lambda_1(\beta,\Pi)}(1+\eta),\notag
	\end{align}
	whenever $n\in[n(\eta),n_b]$,
	where $n_b$ satisfies $b(1+g(b))=\frac{\log n_b}{\Lambda_1(\beta,\Pi)}$. From Lemma \ref{lemmanf}, we know that
	\begin{align}
		n_b=e^{\Lambda_1(\beta,\Pi)b(1+g(b))}=\left(\Pr\left[N_f=1|L_c=b\right]\right)^{-1}.\label{nbexp}
	\end{align}
	2) Note that
	\begin{align}
	&\lim_{n\to\infty}\frac{\log\Pr\left[N_f(b)>n\right]}{n}\notag\\
	&=\max_{l_c}\left\{\lim_{n\to\infty}\frac{\log\left(\Pr\left[L_c(b)=l_c,N_f(b)>n\right]\right)}{n}\right\}\notag\\
	&=\max_{l_c}{\bigg \{}\lim_{n\to\infty}\frac{\log\left(\Pr\left[L_c(b)=l_c\right]\right)}{n},\notag\\
	&\text{ }\text{ }\text{ }\text{ }\text{ }\text{ }\text{ }\text{ }\text{ }\text{ }\text{ }\text{ }\text{ }\text{ }\text{ }\text{ }\text{ }\text{ }\text{ }\text{ }\log\left(\Pr\left[N_f(b)>1|L_c(b)=l_c\right]\right){\bigg \}}\notag\\
	&=\log\left(\Pr\left[N_f(b)>1|L_c(b)=b\right]\right).\notag
	\end{align}
	
	3,4) The proof of 3) and 4) follows by noting that $T_f(b)=N_f(b)L_c(b)$.
	\end{IEEEproof}
	
\subsection{Proof of Theorem \ref{thput}}\label{thputproof}
\begin{IEEEproof}[proof of Theorem \ref{thput}]
	Observe that
	\begin{align}
	\Delta(b)=\lim_{n\to\infty}\frac{\sum_{i=1}^n\beta L_{i}}{n}\frac{n}{\sum_{i=1}^nT_i}=\beta\frac{\mathbb{E}[L_{i}]}{\mathbb{E}[T_i]}.\notag
	\end{align}
From Theorem \ref{fsenm} we know that for a given $\eta>0$, we can find $n(\eta)$ and $b$ large enough such that
	\begin{align}
	\mathbb{E}[T_i]\geq&\int_{n(\eta)b}^{n_bb}\Pr[T_f(b)>t]dt\notag\\
	\geq&\int_{n(\eta)b}^{n_bb}t^{-\frac{\lambda}{\Lambda_1(\beta,\Pi)}(1+\eta)}dt,\notag
	\end{align}
	which, by combing the definition of $n_b$ in Equation (\ref{nbexp}), yields
	\begin{align}
	-\limsup_{b\to\infty}\frac{\log\Delta(b)}{b}=\liminf_{b\to\infty}\frac{\log\mathbb{E}[T_i]}{b}\geq\Lambda_1(\beta,\Pi)-\lambda.\notag
	\end{align}
\end{IEEEproof}
\section{Conclusion}
In this paper, we characterize the delay distribution in a point-to-point
Markovian modulated binary erasure channel with variable codeword
length. Erasure codes are used to encode the information such that a
fixed fraction of bits in the codeword can recover the information. We
use a general coding framework called incremental redundancy
code. In this framework, the codeword is divided into several codeword
trunks and these codeword trunks are transmitted one at a time to the
receiver. Therefore, the receiver gains extra information, which 
is called incremental redundancy, after each transmission.  At the receiver
end, we investigate two different scenarios, namely {\it decoder that
  uses memory} and {\it decoder that does not use memory}. In the
decoder that uses memory case, the decoder caches all previously
successfully transmitted bits. In the decoder that does not use memory
case, received bits are discarded if the corresponding information
cannot be decoded. In both cases, we first assume that the
distribution of codeword length is light-tailed and has an infinite
support. Then, we consider a more realistic case when the codeword
length is upper bounded.

Our results show the following. The transmission delay can be
dramatically reduced by allowing the decoder to use memory. This is
true because the delay is always light-tailed when the decoder uses
memory, while the delay can be heavy-tailed when the decoder does not
use memory. Secondly, analagously to the non-coding case, the tail effect of delay distribution persists
even if the codeword length has a finite support. When the codeword
length is upper bounded, light-tailed delay distribution will turn
into a delay distribution with light-tailed main body whose decay rate
is similar to that of infinite support scenario. Further, we show that the waist of this
main body scales linearly with respect to the increase of maximum
codeword length; heavy-tailed delay distribution will turn into a
delay distribution with heavy-tailed main body, whose waist scales
exponentially with the increase of maximum codeword length. Our
results also provide a benchmark for quantifying the tradeoff between
system complexity (which is determined by code-rate $\beta$, number of
codeword trunks $r$, maximum codeword length $b$ and whether to use
memory at the receiver or not) and the distribution of delay.

\bibliography{IEEEabrv,yangbib}
\end{document}